\documentclass[aps,preprint]{revtex4}
\usepackage{amsfonts}
\usepackage{amsmath}
\usepackage{amssymb}
\usepackage{subfigure}
\usepackage{graphicx}
\usepackage{epstopdf}
\usepackage{epsfig}
\usepackage{float}
\usepackage{color}
\usepackage{ulem}
\usepackage{cancel}
\usepackage{mathrsfs}
\usepackage[colorlinks,linkcolor=blue,citecolor=blue,urlcolor=blue]{hyperref}
\usepackage{url}
\usepackage{cleveref}
\usepackage{ulem}
\usepackage{setspace}

\begin{document}

\preprint{CTP-SCU/2022010}	

\title{Temporal and Spatial Chaos of RN-AdS Black Holes Immersed in Perfect Fluid Dark Matter}
\author{Xingyu Zhou$^{a}$}
\email{zxy_zany@stu.scu.edu.cn}
\author{Yadong Xue$^{a}$}
\email{xueyadong@stu.scu.edu.cn}
\author{Benrong Mu$^{b}$}
\email{benrongmu@cdutcm.edu.cn}
\author{Jun Tao$^{a}$}
\email{taojun@scu.edu.cn}
\affiliation{$^{a}$Center for Theoretical Physics, College of Physics, Sichuan University, Chengdu, 610065, China}
\affiliation{$^{b}$Physics Teaching and Research Section, College of Medical Technology, Chengdu University of Traditional Chinese Medicine, Chengdu 611137, China}

\begin{abstract}
We investigate the thermodynamic chaos of RN-AdS black holes immersed in Perfect Fluid Dark Matter by considering the dynamical equations of the fluid system evolved in the spinodal region. Based on the Melnikov method, it is shown that there exists a critical amplitude that affects the temporal chaos. And the influence of black holes charge and state parameter on the critical amplitude is investigated with specific initial temperature. Then, for inevitable spatial chaos, three different types of portraits are discssued according to the difference between the phase transition pressure and the ambient pressure. Additionally, we check the local equilibrium near saddle points which shows that spatial chaos always exists regardless of the perturbation intensity.
\end{abstract}
\maketitle

\section{Introduction}
Since the prominent contributions of Bekenstein \cite{Bekenstein:1973ur} and Hawking \cite{Hawking:1975vcx}, the thermodynamics of black holes have been extensively studied. Instructively, by regarding the cosmological constant and its conjugate quantity as the thermodynamic pressure and volume respectively, one can investigate black holes in the extended phase space \cite{Kubiznak:2012wp, Kastor:2009wy}. From this point of view, the $P - v$ criticality of RN-AdS black holes in the extended phase space is similar to Van der Waals (VdW) gas/liquid systems \cite{He:2016fiz}. This similarity inspires the investigation for black holes on the chaotic behaviour, which has been widely studied in VdW systems.

As an unpredictable phenomenon for certain dynamical systems in nature, chaos is described with nonlinear or coupled equations for multiple degrees of freedom. Although it is difficult to predict the dynamics state when chaos occurs, there still exist some methods to detect it, such as the Poincare surfaces of section, the Lyapunov characteristic exponents, the Melnikov method and so on \cite{1979A, 1990Poincar, 1981A}. The chaotic motion exists in the VdW fluids system according to the Poincare Melnikov method \cite{Chabab:2018lzf}, which was initially considered in the unstable spinodal region. 

In recent years, the chaos of black holes has been widely studied. For thin rings or discs lying symmetrically around black holes, the growth of chaos in time-like geodesic motion was studied \cite{Semerak:2010lzj, Semerak:2012dx, Sukova:2013jxa}, and chaos in the collapsed polymer model was investigated in \cite{Brustein:2016xzw}. Moreover, it was found that a chaos bound exists in rotating black holes \cite{Jahnke:2019gxr}, and for eccentric black holes binary the chaos order was studied in \cite{Arca-Sedda:2018qgq}. Subsequently, the chaotic behaviors of different types of black holes have been investigated, such as charged AdS black holes \cite{Chabab:2018lzf}, charged Gauss-Bonnet-AdS black holes \cite{Mahish:2019tgv}, Born-Infield-AdS black holes \cite{Chen:2019bwt}, Bardeen AdS black holes with quintessence \cite{Wang:2022oop} and charged dilaton-AdS black holes \cite{Dai:2020wny}. Particularly, chaos and pole-skipping for rotating black holes were investigated in \cite{Blake:2021hjj}, and chaos of charged particles in magnetized Kerr black holes was studied by applying the explicit symplectic integrator \cite{Huang:2022qqu}. Moreover, this topic has been continuously studied by various researchers \cite{Yosifov:2019gwt,Bombelli:1991eg, Chen:2016tmr, DeFalco:2020yys, Hanan:2006uf, Kao:2004qs, Letelier:1996he, Letelier:1999xw, Liu:2018bmn, Polcar:2019kwu, Polcar:2019wfi, Santoprete:2001wz, Wang:2016wcj, Witzany:2015yqa, Slemrod:1985, Wiggins:1990, Wiggins:2003, Tang:2020zhq,Lu:2018mpr,Guo:2015ldd,Guo:2020pgq,Yu:2022ysm,Li:2018wtz}.

According to the standard model of cosmology, dark matter constitutes about $23\%$ of the total mass-energy of the universe \cite{Xu:2017bpz}. Many theoretical models were proposed to be dark matter. In particular, it was introduced by Keislev \cite{Kiselev:2002dx} who described the dark matter as a perfect fluid to explain the asymptotic rotation of spiral galaxies \cite{Guzman:2000zba}. In this way, black holes immersed in Perfect Fluid Dark Matter (PFDM) has been widely investigated \cite{Rahaman:2010xs, Moffat:2004bm, Shaymatov:2020wtj, Zhang:2020mxi,  Das:2020yxw}. For instance, the motion of test particles has been investigated around static black holes \cite{Shaymatov:2020bso} and Bardeen black holes \cite{Narzilloev:2020qtd} immersed in PFDM. The precession frequencies of naked singularities and Kerr-like black holes were also studied under the impact of the PFDM \cite{Rizwan:2018rgs}. Additionally, for rotating black holes immersed in PFDM, the instability, phase transitions \cite{Hendi:2020zyw}, shadow and deflection \cite{Hou:2018avu, Haroon:2018ryd} has been discussed. Moreover, for RN-AdS black holes immersed in PFDM, the Joule-Thomson expansion, thermodynamics and phase transition were also investigated \cite{Cao:2021dcq}. Inspired by these works, we present an analytical study of chaos in PFDM immersed RN-AdS black holes in the extended phase space. 

In this letter, by using the Melnikov functions to calculate the spatially/temporally periodic perturbation, we study the temporal chaos for PFDM immersed RN-AdS black holes in the extended phase space. The influence of PFDM parameter and black holes charge on the chaotic behavior are also discussed. Thereafter, we investigate the different cases of spatial chaos in these black holes in accordance with the difference between the phase transition pressure and the ambient pressure.

The paper is organized as follows. In Section II, we give a brief introduction to the thermodynamic of RN-AdS black holes immersed in PFDM. In Section III, with a spatial periodic perturbation, the thermal chaos was detected on the equilibrium configuration in the Poincare-Melnikov approach, and the influence of PFDM parameter and black holes charge on the temporal chaos was also discussed. In Section IV, we investigate the spatial chaos and check the local properties around the saddle points in three cases. Finally, we conclude our results in Section V.

\section{PFDM immersed RN-AdS black holes in the extended phase space}

We begin with the metric of RN-AdS black holes immersed in PFDM, which takes the form of \cite{Li:2012zx}
\begin{align}
  ds^2=-f(r)dt^2+f(r)^{-1}dr^2+r^2(d \theta^{2}+\sin ^{2} \theta d \phi^{2}),
  \label{metric}
\end{align}
where
\begin{align}
  f(r)=1-\frac{2M}{r}+\frac{Q^2}{r^2}-\frac{\Lambda}{3}r^2+\frac{\lambda}{r}\ln \left( \frac{r}{\left\lvert \lambda \right\rvert}\right),
  \label{fr}
\end{align}
with $Q$, $M$ denoting the charge and mass of black holes respectively, $\lambda$ is the state parameter of PFDM and $\Lambda = -3/\ell^2$ is the cosmological constant with $\ell$ being the AdS radius. In the four-dimensional space-time, the thermodynamic pressure related to $\Lambda $ in the extended phase space yields
\begin{align}
  P=-\frac{\Lambda}{8\pi}.
  \label{P}
\end{align}

Solving $f(r_{+})=0$ to get the event horizon radius $r_+$ and combining the metric function in Eq. (\ref{fr}), we can obtain the mass of black holes
\begin{align}
  M=\frac{4}{3}\pi P r_+^3+\frac{Q^2}{2r_+}+\frac{\lambda}{2}\ln \left(\frac{r_+}{\left\lvert \lambda \right\rvert}\right)+\frac{r_+}{2},
  \label{M}
\end{align}
and the Hawking temperature
\begin{align}
  T=\frac{\lambda}{4\pi r_+^2}+2Pr_++\frac{1}{4\pi r_+}-\frac{Q^2}{4\pi r_+^3}.
  \label{T}
\end{align}

With the specific volume being $v=2r_+$ for the four-dimensional spacetime \cite{Chen:2019bwt}, we could express $P$ in terms of $v$, $T$ as
\begin{align}
  P(v,T)=\frac{T}{v}+\frac{2Q^2}{\pi v^4}-\frac{\lambda}{\pi v^3}-\frac{1}{2\pi v^2}.
  \label{P(v,T)}
\end{align}
\begin{figure}[t]
  \begin{center}
    \renewcommand{\figurename}{Fig}
    \includegraphics[width =8.6cm]{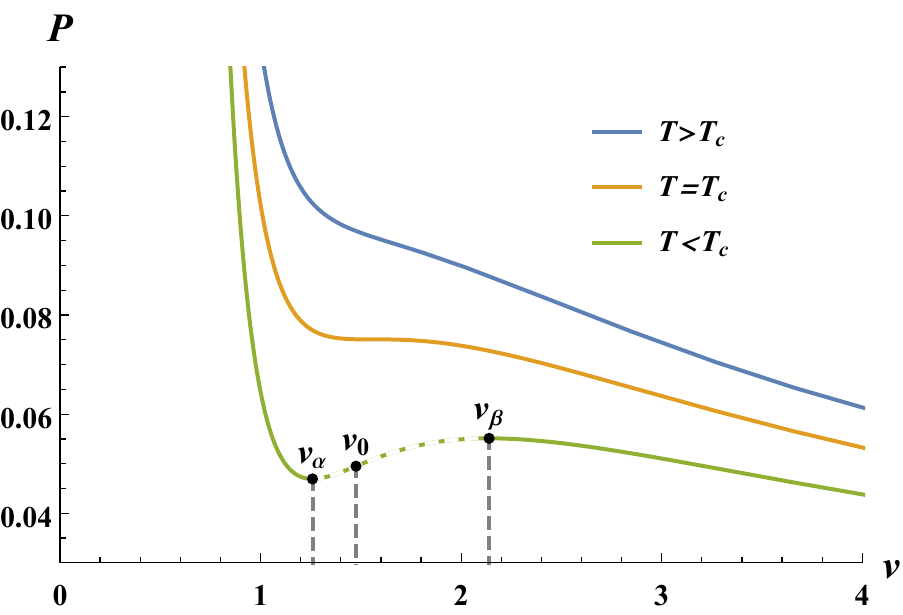}
    \caption{The $P-v$ diagram of RN-AdS black holes immersed in PFDM for $T=0.230$ (Green), $0.268$ (Orange) and $0.300$ (Blue), with $Q=0.7$, $\lambda=1$ and $T_c\simeq 0.268$. For $T<T_c$, the curve could be divided into three regions, i.e. the two green solid lines (with $\frac{\partial P}{\partial v}\Big|_{T}<0$) and one green dashed line (with $\frac{\partial P}{\partial v}\Big|_{T}>0$).}
    \label{figPV}
  \end{center}
\end{figure}

The $P-v$ diagram according to Eq. (\ref{P(v,T)}) are plotted in Fig. \ref{figPV}. When $T<T_{c}$, it could be divided into three regions where the spinodal region is between the extreme point  $v_{\alpha}$ and $v_{\beta}$ of the function \cite{Slemrod:1985}. Especially, we will mainly consider the chaos of black holes near the inflection point $v=v_0$ in the spinodal region defined by $\frac{\partial^2 P}{\partial v^2}\Big|_{T}=0$ where $v_\alpha$ and $v_\beta$ are two extremely points of $P(v,T)$. Analogy with the study of phase transition for the VdW fluid system, we introduce the condition $\frac{\partial P}{\partial v}\Big|_{T=T_c}=\frac{\partial^2 P}{\partial v^2}\Big|_{T=T_c}=0$ and get the critical temperature expressed as \cite{Cao:2021dcq}
\begin{align}
  T_c & =\frac{16 Q^2-3 \lambda  \left(\sqrt{9 \lambda ^2+24 Q^2}-3 \lambda \right)}{\pi  \left(\sqrt{9 \lambda ^2+24 Q^2}-3 \lambda \right)^3}.
  \label{TcL}
\end{align}

\begin{figure}[h] 
  \centering
  \renewcommand{\figurename}{Fig}
  \subfigure[$T_c-\lambda$]{\includegraphics[width=7cm]{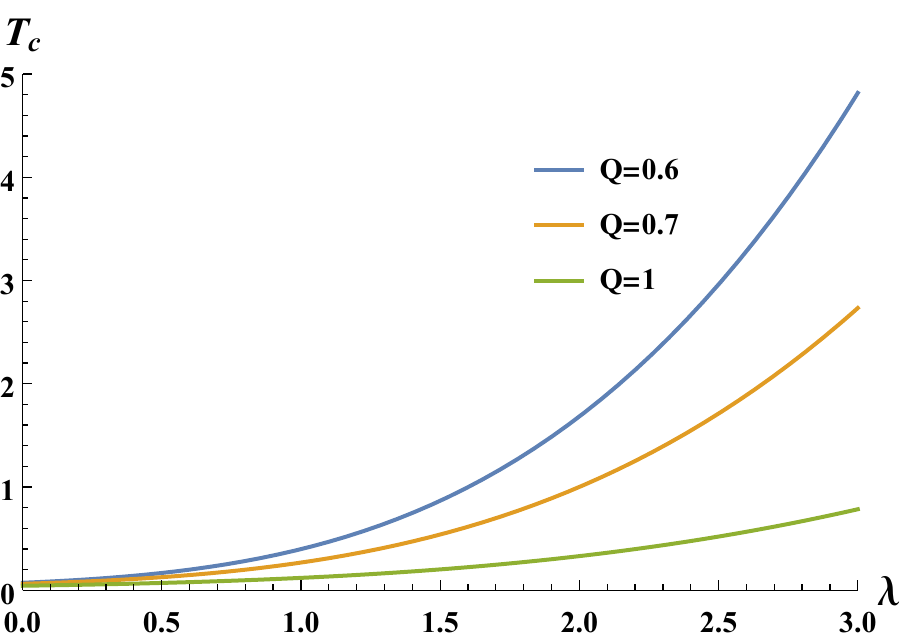}\label{TL}}\quad
  \subfigure[$T_c-Q$]{\includegraphics[width=7cm]{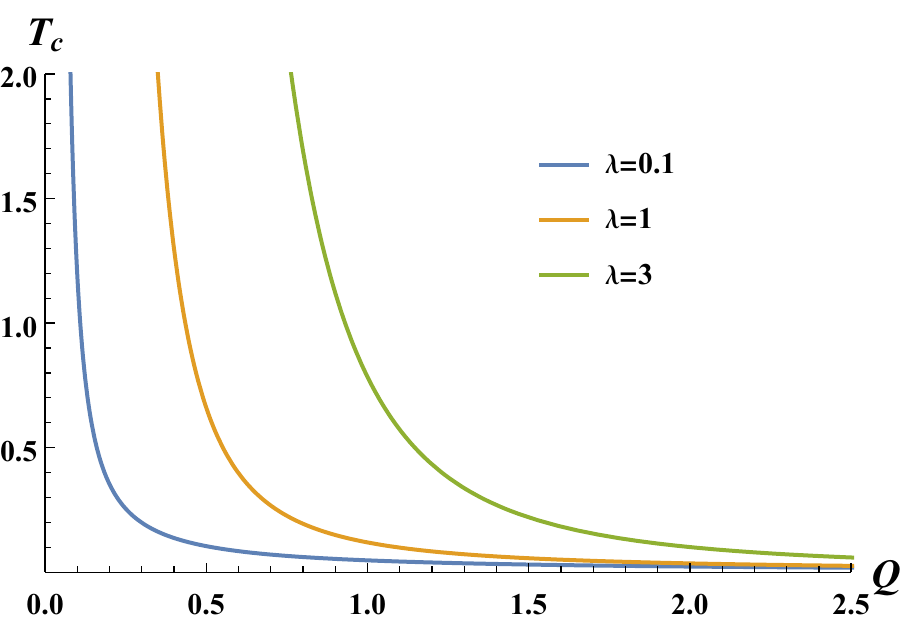}\label{TQ}}
  \caption{Dependence of the critical temperature $T_c$ for RN-AdS black holes with different parameters $\lambda$ and $Q$.}
  \label{TQL}
\end{figure}

It's obviously that $T_c$ is connected to the value of $\lambda$ and $Q$. Just like Fig. \ref{TQL} shows, the critical temperature $T_c$ briefly increase with the increase of $\lambda$ or the decrease of $Q$. 

If we set the condition that $T> T_c$, the dynamic equations considered in next section will reveal a stable homogeneous configuration \cite{Slemrod:1985}. So, basically the condition $T< T_c$ which leads to the unstable spinodal region is essential for the occurrence of chaos. In this way, we will consider the effect of a small periodic fluctuation of the temperature in the next two sections.

\section{Temporal chaos in spinodal region}

In this section, we focus on the temporal chaos for RN-AdS black holes immersed in PFDM in spinodal region. The state equation of black holes as well as the fluid flow's Hamiltonian should be used to construct the dynamical equations and the Melnikov function.

\subsection{Dynamical equations and the Hamiltonian}

We first investigate the dynamical equations of the fluid system at $T_0<T_c$ evolving in the spinodal region under a temporally periodic perturbation, which takes the form as \cite{Slemrod:1985}
\begin{align}
  T=T_0+\varepsilon\gamma \cos(\omega t)\cos M,
  \label{time periodic perturbation}
\end{align}
where $\varepsilon\ll 1$ is a positive perturbation parameter, $\gamma$ represents the amplitude of the perturbation which is relative to the viscosity of the fluid system and $\omega$ is the fluctuation angular frequency.

We assume the PFDM immersed RN-AdS black holes flow moves along $x$-axis in a constant volume tube of unit cross section, and the mass of the column between $x_0$ and $x$ at time $t$  is given by \cite{Chabab:2018lzf,Slemrod:1985}

\begin{align}
  M=\int^{x}_{x_0}\rho(\xi,t)d\xi,
  \label{M(x,t)}
\end{align}
with $\rho(\xi,t)$ being the density of the unit tube on $x=\xi$ at time $t$. Express $x$ with $M$ and $t$ as $x=x(M,t)$, the specific volume $v$ and the velocity $u$ could be defined as
\begin{align}
  v(M,t) \equiv x_M(M,t),\quad u(M,t)\equiv x_t(M,t),
\end{align}
where a subscript stands for the derivative with respect to the index.

We could rewrite the system equations in Lagrangian coordinates as $v_t=u_M$ and $u_t=\tau_M$ where $\tau$ denoting the Piola stress tensor defined in Korteweig's theory. The fluid was considered as thermoelastic, slightly viscous and isotropic within an additional stress component given by the Korteweg theory of capillarity \cite{Slemrod:1985}. And the Piola stress takes the form as
\begin{align}
  \tau=-P(v,T)+\mu u_M-Av_{MM},
  \label{Piola stress}
\end{align}
where $A$ is a positive constant, and the viscosity $\mu=\varepsilon \mu _0$ denotes a small positive constant. To simplify the fourier series expansion, we could introduce a set of variables $\tilde{M}=sM$, $\tilde{t}=st$, $\tilde{x}=sx$ with $s\equiv 2\pi/M$, then the ranges of $M$, $t$ and $x$ are all in $[0,2\pi]$. For the dynamical equation of the black hole, we can write the Hamiltonian as \cite{Chabab:2018lzf}
\begin{align}
  H(x,u;T)=\frac{1}{\pi}\int^{2\pi}_{0}\left[\frac{u^2}{2}-\int^{v}_{v_0}P(v,T)dv+\frac{As^2}{2}v^2_{M}\right]dM.
  \label{23}
\end{align}

To obtain the Hamiltonian at the thermodynamic equilibrium in the frist two modes for simplify, one can perform a expansion of $P(v,T)$ for $v$ and $T$ around $(v_0,T_0)$ up to $\mathcal{O} (v^4)$, then expand $v$, $u$ in the Fourier series and take terms of $\cos M$ and $\sin M$ with $M\in [0,2\pi]$. The frist two modes of Hamiltonian in Eq. (\ref{23}) can be expressed as
\begin{equation}
\begin{aligned}
  H_2(x,u)=&\frac{u_1^2+u_2^2}{2}-\frac{\bar{P}_v}{2}(x_1^2+x_2^2)+\frac{As^2}{2}(x_1^2+4x_2^2)-\frac{\bar{P}_{vvv}}{32}(x_1^4+4x_1^2x_2^2+x_2^4)\\&-
  \frac{1}{8v_0}\varepsilon\gamma\cos (\omega t)x_1\left[8-\frac{4}{v_0}x_2+\frac{2}{v_0^2}(x_1^2+2x_2^2)\right],
  \label{14}
\end{aligned}
\end{equation}
where the higher mode terms vanish at $x_n=u_n=0$ for $n\geq3$ \cite{Slemrod:1985}, and
\begin{align}
  \bar{P}_v&=\frac{-8Q^2+3\lambda v_0+v_0^2-\pi T_0v_0^3}{\pi v_0^5},\\\bar{P}_{vvv}&=-6\frac{40Q^2-10\lambda v_0-2v_0^2+\pi T_0v_0^3}{\pi v_0^7},
\end{align}
while we define the value of $P(v,T)$ at $(v_0,T_0)$ as $\bar{P}\equiv P(v_0,T_0)$.

From the Hamiltonian, we could simply derive the corresponding equations of motion as follows
\begin{equation}
  \begin{aligned}
      \dot{x}_1=&u_1,\\
      \dot{x}_2=&u_2,\\
      \dot{u}_1=&(\bar{P}_v-As^2)x_1+\frac{\bar{P}_{vvv}}{8}(x_1^3+2x_1x_2^2)-\varepsilon\mu_0su_1\\&+\frac{\varepsilon\gamma}{v_0}\cos (\omega t)\left(1+\frac{3}{4v_0^2}x_1^2-\frac{1}{2v_0}x_2+\frac{1}{2v_0^2}x_2^2\right),\\
      \dot{u}_2=&(\bar{P}_v-4As^2)x_2+\frac{\bar{P}_{vvv}}{8}(x_2^3+2x_2x_1^2)-4\varepsilon\mu_0su_2\\&+\frac{1}{2v_0^2}\varepsilon\gamma\cos (\omega t)x_1\left(\frac{2}{v_0}x_2-1\right).
    \label{17}
  \end{aligned}
\end{equation}

 By solving these equations, we could get the Melnikov function of temporal chaos for PFDM immersed RN-AdS black holes.

\subsection{Temporal chaos of RN-AdS black holes immersed in PFDM}

Before we solve Eq. (\ref{17}) to study the chaos of these black holes, we should research the properties of these equations. Firstly, we linearize this differential equations by considering $\gamma=0$ and abandoning the higher modes, such that we could obtain the linear equation as 
\begin{equation}
  \dot{\mathbf{z}}_L(t)=L\mathbf{z}_L(t),
\end{equation}
where $\mathbf{z}_L=(x_1,x_2,u_1,u_2)^T$ and the matrix $L$ is given by
\begin{equation}
  L=  \begin{pmatrix}
    0&0&1&0 \\
    0&0&0&1 \\
    -As^2+\bar{P}_v&0&-\varepsilon\mu_0s&0\\
    0&-4As^2+\bar{P}_v&0&-4\varepsilon\mu_0s
  \end{pmatrix}.
\end{equation}
Then we can get the eigenvalues of $L$ as
\begin{equation}
  \begin{aligned}
    \lambda_L^{1,2}&=-\frac{\varepsilon\mu_0s}{2}\pm\sqrt{\bar{P}_v-As^2},\\
    \lambda_L^{3,4}&=-2\varepsilon\mu_0s\pm\sqrt{\bar{P}_v-4As^2},
  \end{aligned}
\end{equation}
where we omit the higher orders of $\varepsilon$. 

When $\bar{P}_v/(4A)<s^2<\bar{P}_v/A$, the first mode is unstable as $\lambda^1_L>0$ and $\lambda^2_L<0$, which implies a saddle point. And it's unstable for higher modes while the real part of $\lambda_L^{3,4}$ is both negative \cite{Mahish:2019tgv}.

\begin{figure}[]
  \begin{center}
    \renewcommand{\figurename}{Fig}
    \includegraphics[width = .6\textwidth]{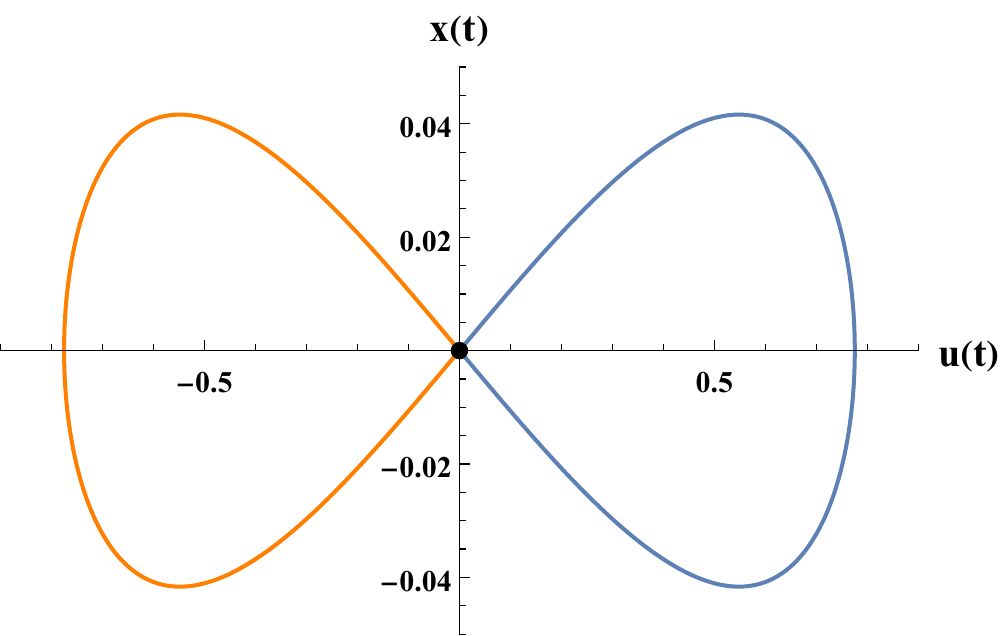}
    \caption{Homoclinic orbit of the unperturbed equations with $T_0 = 0.9T_c$, $Q=0.7$, $\lambda=1$, $A=0.2$ and $s=0.001$.}
    \label{figHorbit}
  \end{center}
\end{figure}

To solve these partial differential equations with perturbation, we rewrite Eq. (\ref{17}) in phase space as
\begin{align}
  \dot{\mathbf{z}}=f(\mathbf{z})+\varepsilon g(\mathbf{z},t),
  \label{dynamical equation}
\end{align}
here $\varepsilon\ll 1$ being a small perturbation parameter and $g(\mathbf{z},t)$ being the periodic function of $t$. Considering the unperturbed system with $\varepsilon=0$ and $\dot{\mathbf{z}}=f(\mathbf{z})$, the Hamiltonian would generate a homoclinic orbit with a fixed point which has smooth flow and conserved energy. It allows a solution $\mathbf{z}=\left\lvert \mathbf{z}_0(t) \right\rvert$, which can be expressed as \cite{Slemrod:1985}
\begin{equation}
  \mathbf{z}_0(t)=\begin{pmatrix}
    \frac{4 a }{\sqrt{-\bar{P}_{vvv}}}\text{sech}(a t)\\
    0\\
    \frac{-4 a^2 }{\sqrt{-\bar{P}_{vvv}}}\text{sech}(a t)\tanh(at)\\
    0
  \end{pmatrix},
\end{equation} 
where $a^2\equiv \bar{P}_v-As^2$. And it forms a typically homoclinic orbit shown in Fig. \ref{figHorbit}. We can see that the stable and unstable manifolds intersects themselves at the saddle equilibrium point located at the original point.

With the temporal perturbation to the system, the stable and unstable manifolds will intersect in a complex way, resulting in countless intersections in the process of reaching equilibrium. Parameterizing the unperturbed homoclinic orbit by time, we consider the perpendicular distance from the stable to unstable manifolds in Poincare surfaces when $t=t_0$ \cite{Wiggins:1990},
\begin{align}
  d(t_0)=\frac{\varepsilon M(t_0)}{\left\lvert f(\mathbf{z}_0(0)) \right\rvert},
  \label{eq.12}
\end{align}
where $M(t_0)$ is the Melnikov function, which takes the form as \cite{1979A, 1990Poincar, 1981A}
\begin{align}
  M(t_0)=\int^{+\infty}_{-\infty}f^T(\mathbf{z}_0(t-t_0))\Omega_ng(\mathbf{z}_0(t-t_0))dt,
  \label{eq.13}
\end{align}
and the subscript $n$ stands for the freedom degrees of perturbations. When $n=2$, $\Omega_{n}$ takes the form as
\begin{align}
  \Omega_{2}=
  \begin{pmatrix}
    0  & 1 & 0 & 0\\
    -1 & 0 & 0 & 0\\
    0 & 0 & 0 & 1\\
    0 & 0 & -1 & 0
  \end{pmatrix}.
  \label{eq.14}
\end{align}

The Melnikov function in Eq. (\ref{eq.12}) can estimate the distance for the transverse intersections of stable and unstable orbits. If the zero point $M(t_{0})=0$ exists after considering suitable value of perturbation $\varepsilon>0$, it means that the two kinds of manifolds will intersect transversely at $t_0$ in the Hamiltonian system \cite{Polcar:2019kwu, Chabab:2018lzf, 1979A, 1990Poincar, 1981A}. It represents the Poincare map has a Smale horseshoe(a invariant hyperbolic set) for such intersecting orbits, and indicates the existence of chaos. Then, we need to solve the equation and check whether $M(t_0)$ crosses zero \cite{1990Poincar}.

With the PFDM immersed RN-AdS black holes quenched to the spinodal region, we can study the effect of a weak temporally periodic perturbation in this subsection.  With the dynamical equations for the fluid flow in Eq. (\ref{17}), if we consider $\mathbf{z}=\left(x_1,u_1,x_2,u_2\right)^T$, the functions in Eq. (\ref{dynamical equation}) can be expressed as
\begin{align}
  f(\mathbf{z}_0(t'))= \begin{pmatrix}\frac{4 a^2 }{\sqrt{-\bar{P}_{vvv}}}\text{sech}\left(at'\right)\tanh(a(t'))\\
  \frac{4a^3}{\sqrt{-\bar{P}_{vvv}}}\text{sech}\left(at'\right)+\frac{8a^3}{\sqrt{-\bar{P}_{vvv}}}\text{sech}^3\left(at'\right)\\
  0\\
  0
\end{pmatrix},
\end{align}
and
\begin{align}
  g(\mathbf{z}_0(t'))=\begin{pmatrix}
    0\\
    \frac{\gamma}{v_0}\cos(\omega t)-4a^2\eta\text{sech}^2\left(at'\right)\\
    0\\
    -\frac{2a\gamma}{v_0^2\sqrt{-\bar{P}_{vvv}}}\cos(\omega t)\text{sech}\left(at'\right)
  \end{pmatrix},
\end{align}
where  we set $t'\equiv t-t_0$ and
\begin{equation}
  \eta=\frac{3\gamma}{v_0^3\bar{P}_{vvv}}\cos(\omega t)+\frac{s\mu_0}{\sqrt{-\bar{P}_{vvv}}}\sinh(at').
\end{equation}

For the orbit reaching equilibrium, the Melnikov function can be expressed as
\begin{align}
  M(t_0)=\int^{+\infty}_{-\infty}\left[\theta\;\text{sech}^2(a t) \tanh ^2(a t)+\alpha \;\text{sech}(a t) \tanh (a t)  \left(1-\beta  \text{sech}^2(a t)\right) \cos (\omega t +\omega t_0)\right]dt,
  \label{33}
\end{align}
where we use $t$ instead of $t'$ for simplicity, and
\begin{align}
  \theta =\frac{16 s \mu_0  a^4}{\bar{P}_{vvv}},\quad \alpha=-\frac{4\gamma v_0}{a^2\sqrt{-\bar{P}_{vvv}}},\quad \beta=\frac{12a^2}{v_0^2\bar{P}_{vvv}}.
  \label{35}
\end{align}

We can get the specific expression of the Melnikov function as
\begin{align}
  M(t_0)=\gamma \omega K\sin(\omega t_0)+\mu_0sL,
\end{align}
with
\begin{align}
  K=-\frac{4\pi}{v_0\sqrt{-\bar{P}_{vvv}}}\left[1-\frac{2}{v_0^2\bar{P}_{vvv}}(\omega^2+a^2)\right]\text{sech}(\frac{\pi \omega}{2a}),\quad L=\frac{32a^3}{3\bar{P}_{vvv}}.
\end{align}

To ensure the Melnikov function has zero point and allow the existence of chaos after a weak and temporally periodic perturbation, we require $\left\lvert\frac{s\mu_0 L}{\gamma \omega K}\right\lvert\leq 1$ and then obtain a critical quantity $\gamma_c$ as the boundary of chaos which takes the form as \cite{Chabab:2018lzf}
\begin{align}
  \gamma_c=\left\lvert\frac{s\mu_0 L}{\omega K}\right\lvert.
\end{align}

The dependences of $\gamma_c$ on the initial temperature, the PFDM parameter and the charges of the black holes are worth investigating in detail in this subsection. As we could found that the $\gamma_c$ is influenced by $T$, $\lambda$ and $Q$. 

\subsection{Chaotic region of temporal chaos}

It should be noticed that only when $T<T_c$ that we could get the spinodal region existed, which is the essential condition of chaos. As $T_c$ was calculated from $Q$ and $\lambda$ in Eq. (\ref{TcL}), the selection of initial temperature $T_0$ is very important. Such that, we could involve two kinds of value choices of $T_0$, where $T_0$ is proportional to $T_c$ or has a fixed value, which result in some meaningful diagrams as follows.

\begin{figure}[] 
  \centering
  \renewcommand{\figurename}{Fig}
  \subfigure[$\lambda=1$]{\includegraphics[width=7cm]{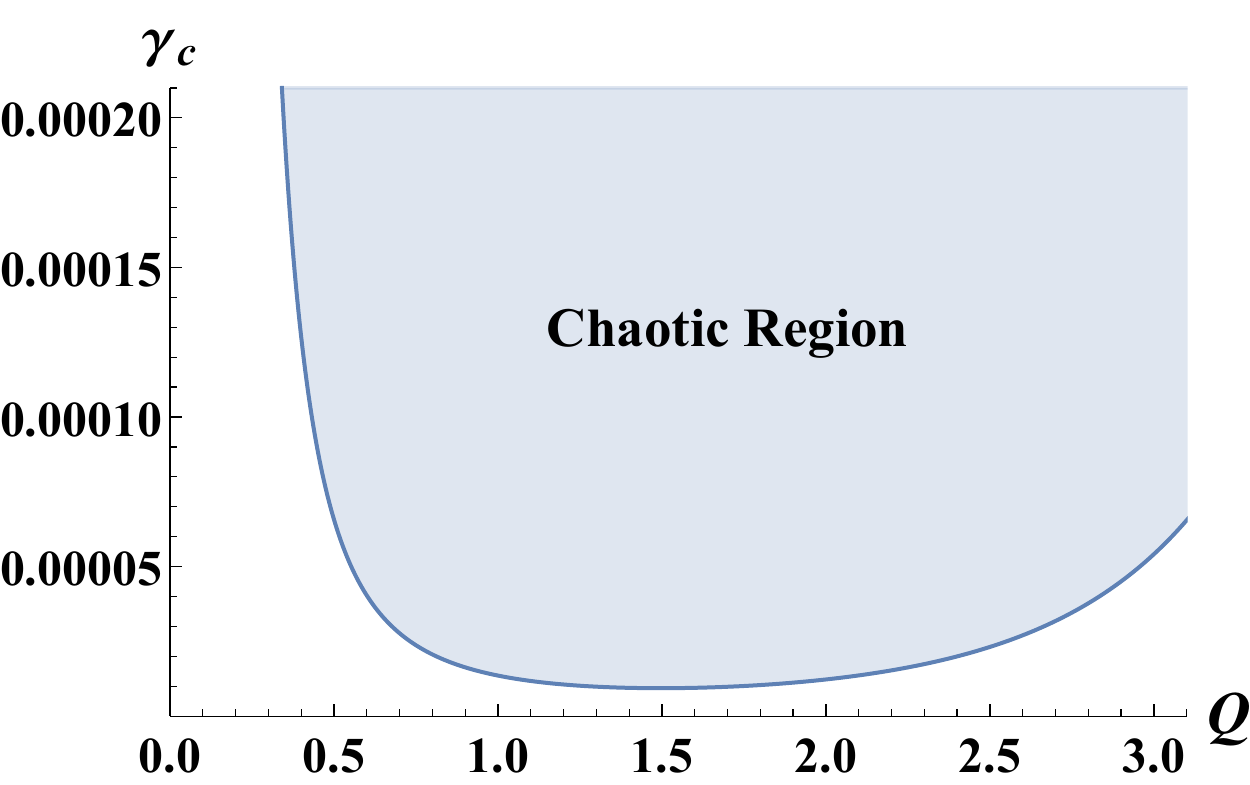}\label{Q1}}\quad
  \subfigure[$Q=0.7$]{\includegraphics[width=7cm]{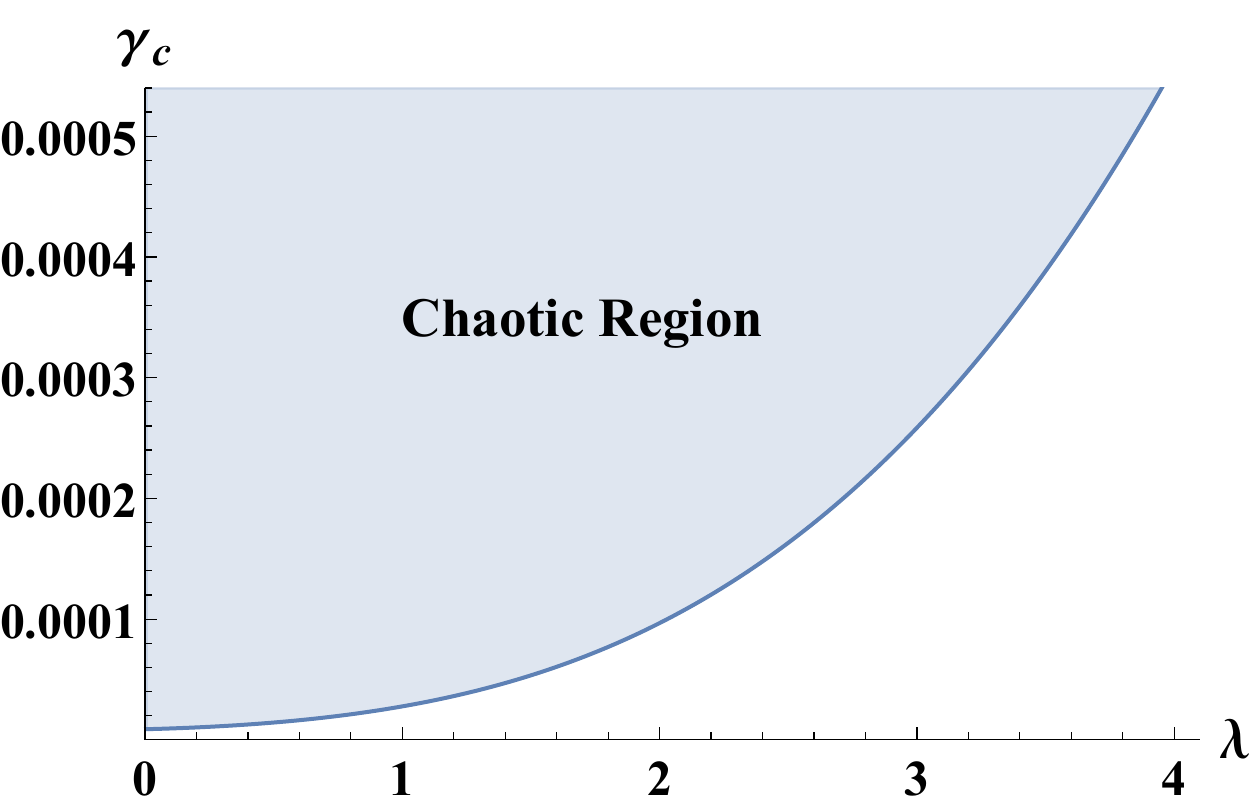}\label{L1}}
  \caption{Dependence of the critical amplitude $\gamma_c$ on the charge $Q$ (left) and the PFDM parameter $\lambda$ (right) under the condition of $T_0=0.9 T_c$. The shaded region denotes the existence of chaos. We set $\lambda=1$ for Fig. \ref{Q1} and $Q=0.7$ for Fig. \ref{L1}.}
  \label{gamac}
\end{figure}

Consider $T_0=0.9T_c$ first, the critical amplitude $\gamma_c$ are plotted with fixed $\lambda=1$ or $Q=0.7$ in Fig. \ref{gamac} respectively, where the shaded region denotes the existence of chaos. With $\lambda=1$ in Fig. \ref{Q1}, we find that it is easier to produce temporal chaos black holes with a small charge, but $\gamma_c$ diverges when the charge of the black holes approaches zero, which indicates that it is hard for AdS black holes immersed in PFDM to produce temporal chaos. While for $Q=0.7$ fixed in Fig. \ref{L1}, the critical amplitude is a monotonically increasing function of $\lambda$ which shows that the addition of PFDM makes the occurrence of chaos more difficult.

\begin{figure}[b] 
  \centering
  \renewcommand{\figurename}{Fig}
   \subfigure[$\lambda=1$]{\includegraphics[width=7cm]{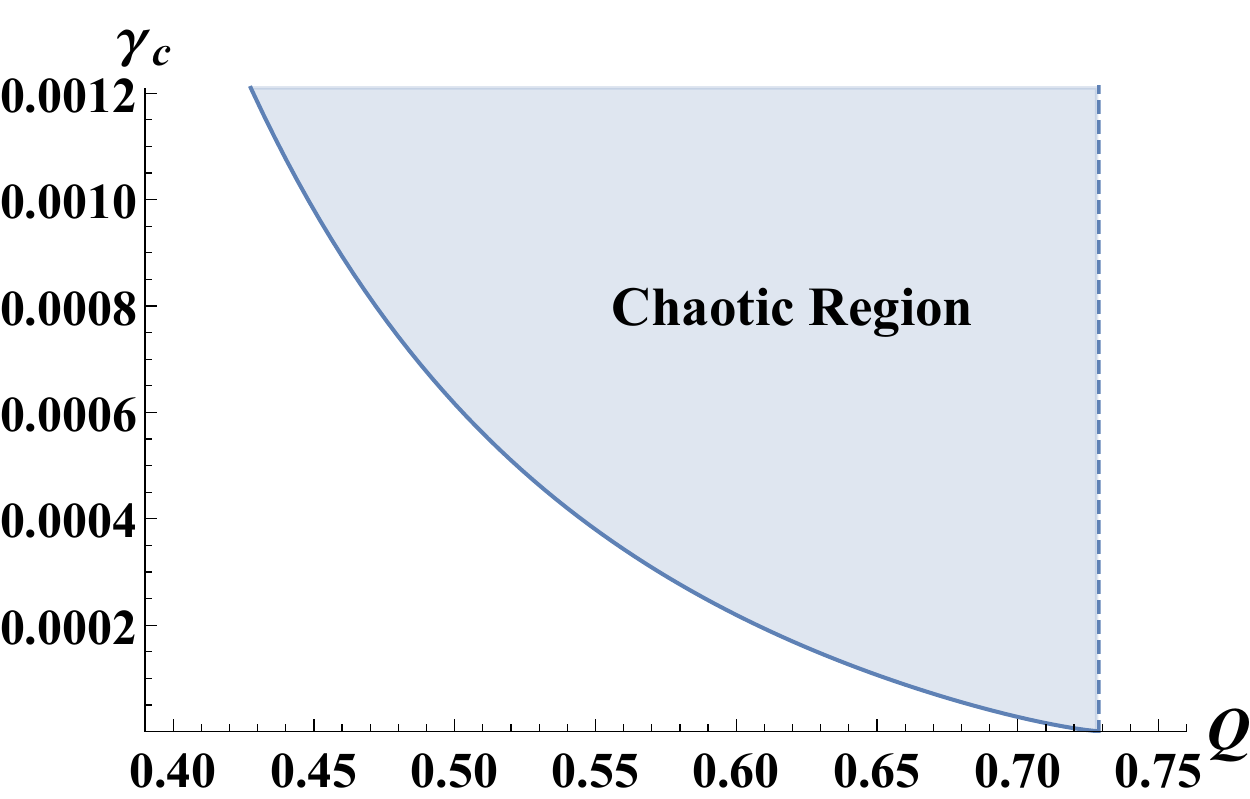}\label{Q2}}\quad
  \subfigure[$Q=0.7$]{\includegraphics[width=7cm]{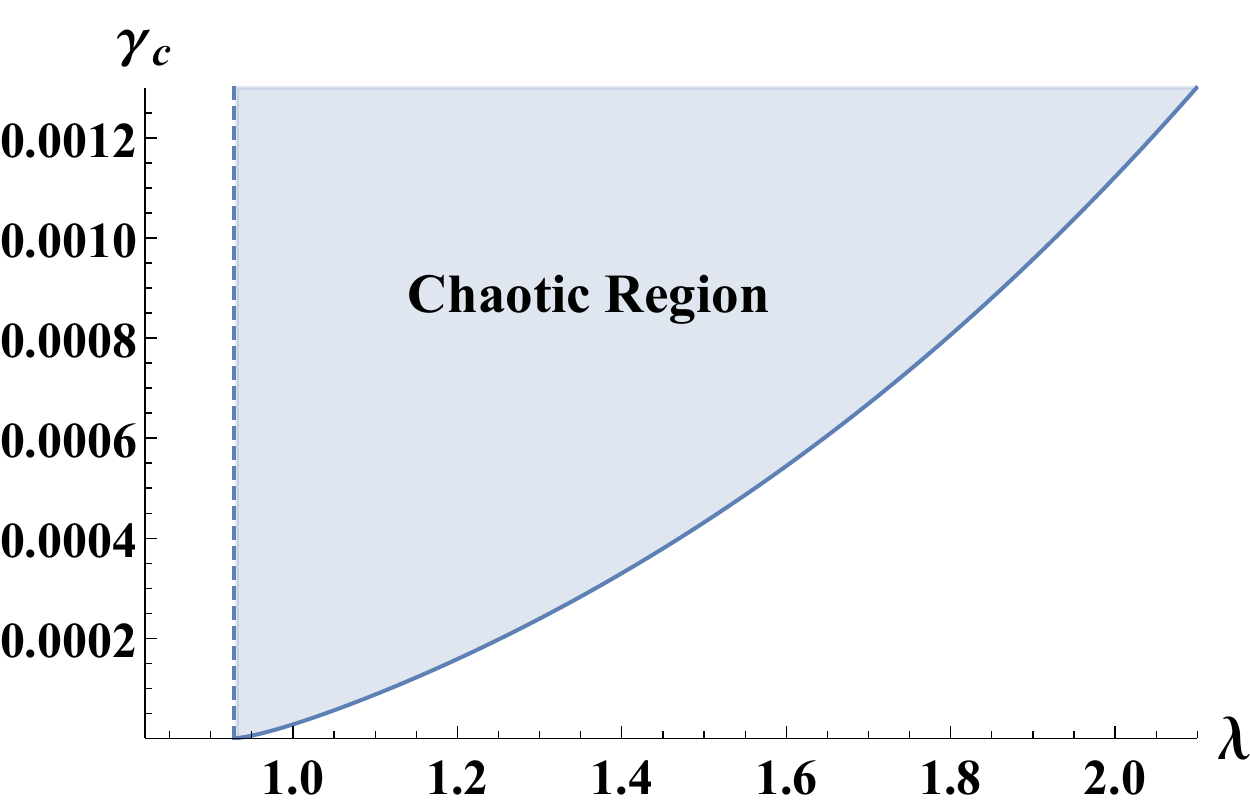}\label{L2}}
  \caption{Dependence of the critical amplitude $\gamma_c$ on the charge $Q$ (left) and the PFDM parameter $\lambda$ (right) under the condition of $T_0=0.240993$. The shaded region denotes the existence of chaos. We set $\lambda=1$ (left) and $Q=0.7$ (right).}
  \label{figgamac}
\end{figure}

When setting the initial temperature to a certain value $T_0=0.240993$, we plot the critical amplitude $\gamma_c$ with fixed $\lambda=1$ or $Q=0.7$ in Fig. \ref{figgamac} respectively. It's obvious that, different from $T_0=0.9T_c$ which holds $T_0<T_c$ for any parameters allowed, $T_0$ with fixed value indicates the boundary where $T_0=T_c$. Just as Fig. \ref{TQL} shows, with fixed $\lambda$ , the critical temperature $T_c$ decrease with larger $Q$. Such that it's easy to illustrate that this diagram Fig. \ref{Q2} should have a upper bound of charge $Q$ accord with what we have calculated.  In a similar way, we could easily demonstrate that Fig. \ref{L2} should have a lower bound of $\lambda$.

\begin{figure}[H]
    \centering
    \renewcommand{\figurename}{Fig}
    \subfigure[$\gamma=0.0000276<\gamma_c$]{\includegraphics[width=7cm]{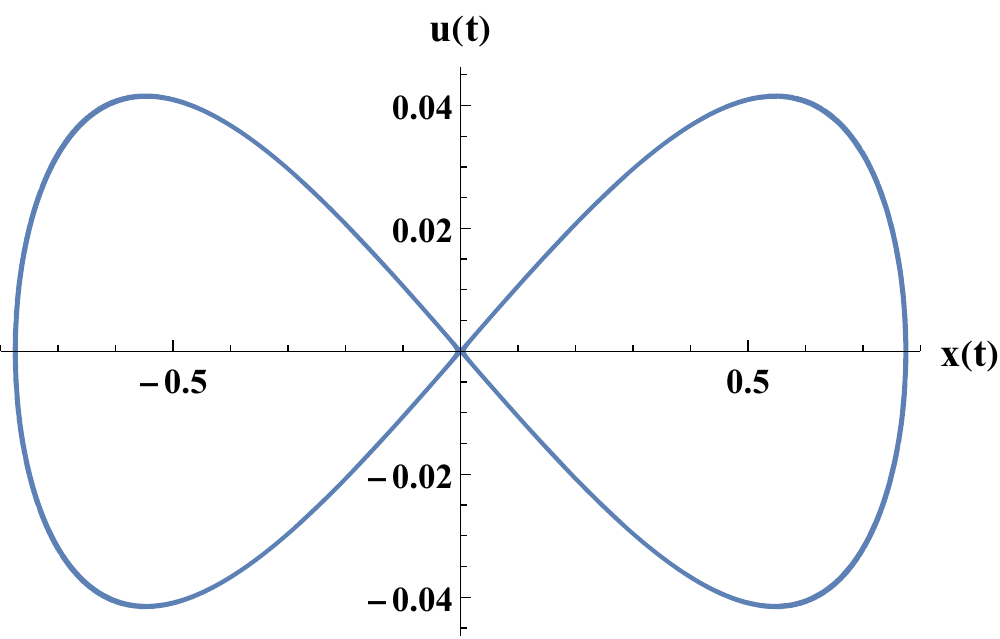}\label{st4}}\quad
    \subfigure[$\gamma=5>\gamma_c$]{\includegraphics[width=7cm]{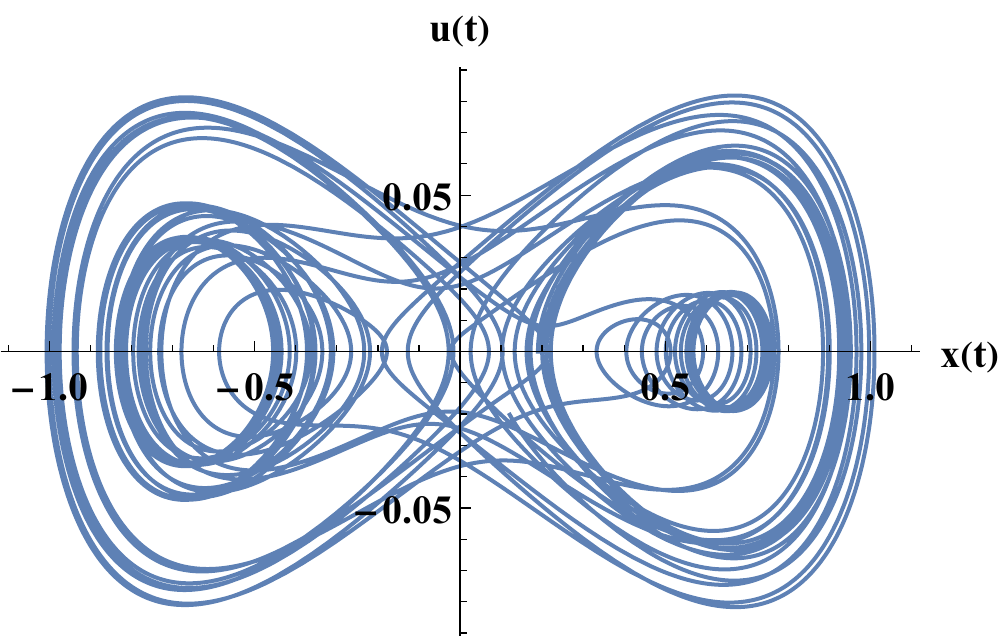}\label{st5}}
    \caption{The unperturbed evolution orbit (left) and perturbed evolution orbit (right) for the critical value $\gamma_c \approx 0.0000277$ in $x-u$ plane with $T_0 = 0.9T_c$. We set $\omega = 0.01$, $\varepsilon = 0.001$, $\mu_0 = 0.1$, $Q=0.7$, $\lambda=1$, $A=0.2$ and $s=0.001$. }
    \label{figchaos}
\end{figure}

The unperturbed and perturbed dynamic systems are solved numerically and shown in Fig. \ref{figchaos} respectively. If $\gamma=0.0000276< \gamma_c$, the homoclinic orbit maintains the original evolution trend on the phase panel as in Fig. \ref{st4} which shows predictable nature. But when $\gamma=5>\gamma_c$, the evolution orbit turns out to be unpredictable and extremely complex as shown in Fig. \ref{st5}, which marks the emergence of chaos in the dynamic system.

In general, the amplitude of the perturbation $\gamma$ determines the occurrence of the temporal chaos and there also exist a critical amplitude $\gamma_c$ as the lower bound for RN-AdS black holes immersed in PFDM. Our results show that, the charge $Q$ and the PFDM state parameter $\lambda$ of black holes affect the bound of temporal chaos.

\section{Spatial chaos in the equilibrium configuration}
In this section, we add a tiny spatially thermal periodic perturbations in the equilibrium state of the PFDM immersed RN-AdS black holes at temperature $T_0 < T_c$. The sub-critical temperature is given as follows \cite{Slemrod:1985},
\begin{align}
  T=T_0+\varepsilon\cos px,
  \label{eq38}
\end{align}
here $p$ is the fluctuation angular frequency. According to the VdW-Korteweg theory, the Piola stress tensor in Eq. (\ref{Piola stress}) can be written as
\begin{align}
  \tau=-P(v,T)-Av_{xx},
  \label{eq.21}
\end{align}
where $A$ is a positive constant and $P(v,T)$ satisfies Eq. (\ref{P(v,T)}). We obtain $\tau_x=0$ for static equilibrium without external forces, so that $\tau=\text{const.}=-B$, with $B$ being the ambient pressure as $\left\lvert x\right\rvert \rightarrow \infty$ \cite{Mahish:2019tgv}. In this way, we can obtain 
\begin{align}
  Av_{xx}+P(v,T)=B.
  \label{eq40}
\end{align}

For the unperturbed system with any temperature below critical value, the nonlinear systems in above equation have three fixed points denoted as $v_1$, $v_2$ and $v_3$ respectively.
After adding a small spatial perturbation as in Eq. (\ref{eq38}), we could have
\begin{align}
  Av_{xx}+P(v,T_0)+\frac{\varepsilon\cos px}{v}=B.
  \label{dynamics temporal}
\end{align}

\begin{figure}[b]
  \centering
  \renewcommand{\figurename}{Fig}
  \subfigure[$P-v$]{\includegraphics[width=7cm]{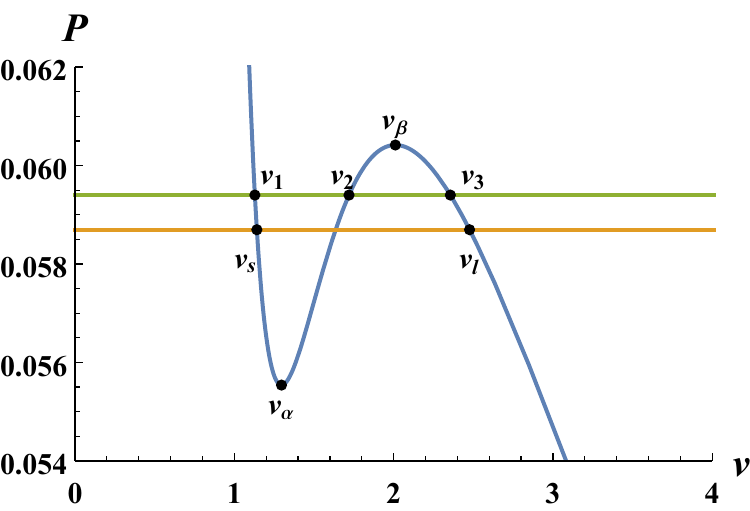}\label{1pv}}\quad
  \subfigure[$v_x-v$]{\includegraphics[width=7cm]{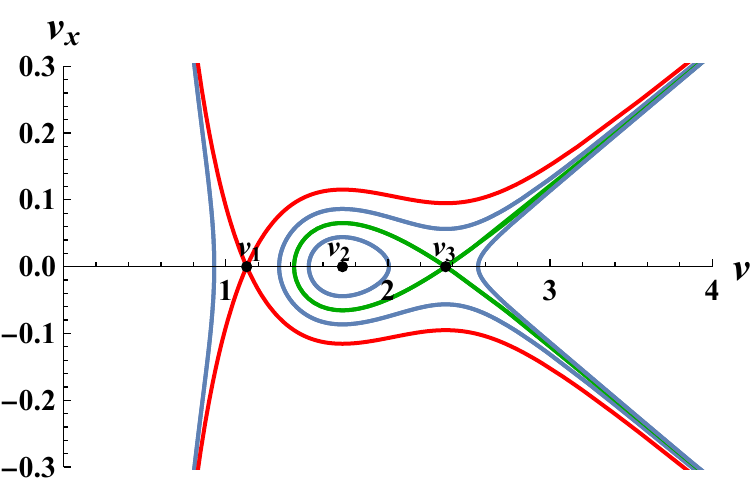}\label{1vxv}}
  \caption{$P-v$ isotherm (left) and the corresponding unperturbed phase portrait (right) in $v_x-v$
    plane for Case 1 with $T_0=0.9T_c$, $Q=0.7$ and $\lambda=1$.}
    \label{f7}
\end{figure}

The Maxwell equal area construction taking the form as
\begin{equation}
  \int^{v_l}_{v_s}(P(v,T_0)-P_0)dv=0,
  \label{phase transition pressure}
\end{equation}
here $v_l$ and $v_s$ are shown in Fig. \ref{1pv}. When set $\lambda=1$, $Q=0.7$ and $T_0=0.9T_c$, we can obtain the phase transition pressure $P_0=0.0587$.

Comparing the value of the phase transition pressure $P_0$ and the ambient pressure $B$, we can generate three different types of portraits in the $v_x-v$ phase plane:

\begin{spacing}{2.0}
\noindent{\textbf{A. Case 1}}
\end{spacing}

In this case, the ambient pressure is in the range $P_0<B<P(v_\beta,T_0)$. Setting $B=0.0594>P_0$, the $P-v$ isotherm and the corresponding unperturbed phase portrait are plotted in Figs. \ref{1pv} and \ref{1vxv} respectively. For the $P-v$ isotherm, the green line reprensents $P=B$ while the orange line represents $P=P_{0}$. For the corresponding unperturbed $v_{x}-v$ phase portrait, there is a homoclinic orbit represented by a green curve connecting one saddle point $v_3$ to itself and $v_2$ is the center of orbit. The red lines pass through $v_1$ are not homoclinic or heteroclinic orbits. And the blue auxiliary lines show the intermediate state of the orbits in all these three cases.

\begin{spacing}{2.0}
\noindent{\textbf{B. Case 2}}
\end{spacing}

\begin{figure}[H]
  \centering
  \renewcommand{\figurename}{Fig}
  \subfigure[$P-v$]{\includegraphics[width=7cm]{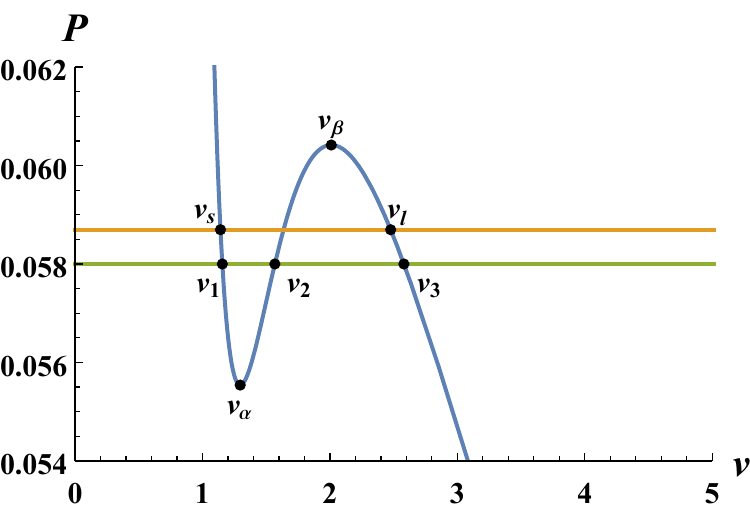}\label{2pv}}\quad
  \subfigure[$v_x-v$]{\includegraphics[width=7cm]{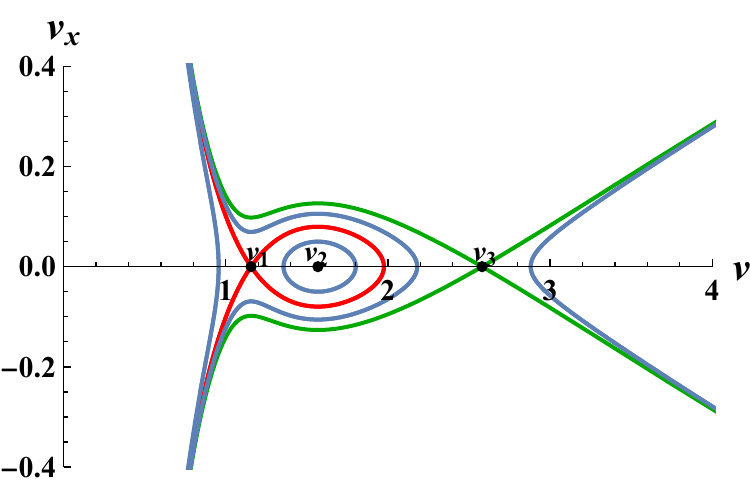}\label{2vxv}}
  \caption{$P-v$ isotherm (left) and the corresponding unperturbed phase portrait (right) in $v_x-v$
    plane for Case 2. And we also fixed $T_0=0.9T_c$, $Q=0.7$ and $\lambda=1$.}
    \label{f8}
\end{figure}

Choosing the range of the ambient pressure in $P(v_\alpha,T_0)<B<P_0$, and set $B=0.0580$, the $P-v$ isotherm and the corresponding unperturbed phase portrait are plotted in Figs. \ref{2pv} and \ref{2vxv} respectively. In this case, we can get the homoclinic orbit represented by the red line connecting a saddle point $v_1$ to itself . Same as the case above, $v_2$ is the center of the homoclinic orbit. 

\begin{spacing}{2.0}
\noindent{\textbf{C. Case 3}}
\end{spacing}

\begin{figure}[H]
  \centering
  \renewcommand{\figurename}{Fig}
  \subfigure[$P-v$]{\includegraphics[width=7cm]{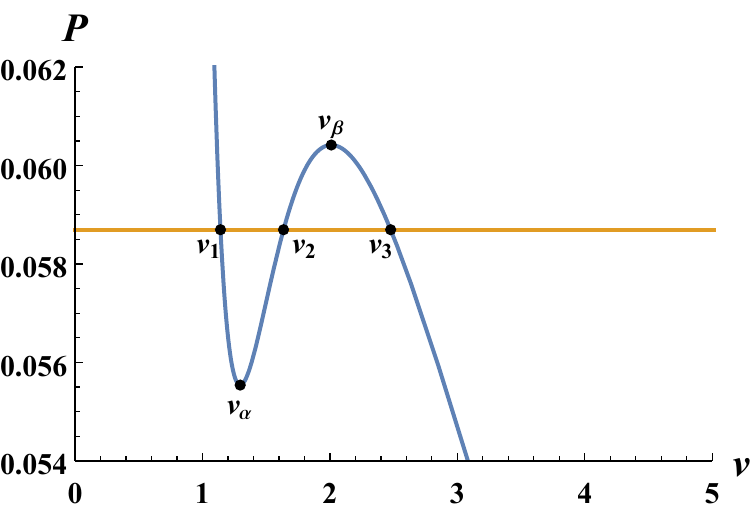}\label{3pv}}\quad
  \subfigure[$v_x-v$]{\includegraphics[width=7cm]{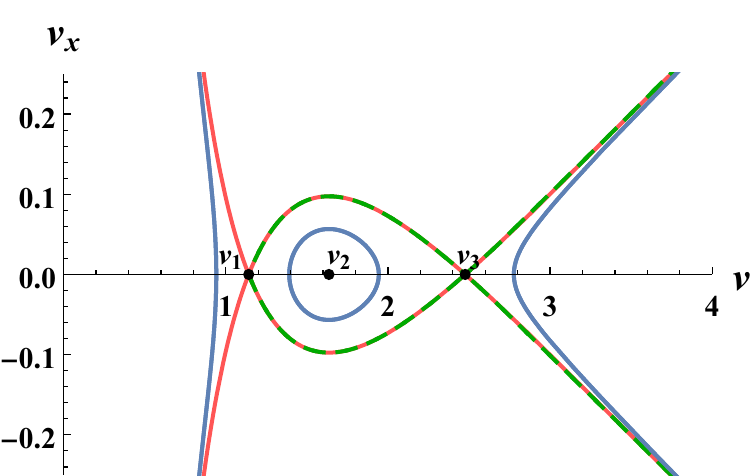}\label{3vxv}}
  \caption{$P-v$ isotherm (left) and the corresponding unperturbed phase portrait (right) in $v_x-v$
    plane for Case 3 where $T_0=0.9T_c$, $Q=0.7$ and $\lambda=1$.}
    \label{f9}
\end{figure}

In this case the ambient pressure $B=0.0587=P_0$, the $P-v$ isotherm and the corresponding unperturbed phase portrait are plotted in Figs. \ref{3pv} and \ref{3vxv} respectively. In Fig. \ref{3pv}, the orange line reprensents $P=B=P_0$ which results in a heteroclinic orbit, the green line, connecting $v_1$ with $v_3$ in Fig. \ref{3vxv}. Similar to the case we have mentioned, $v_2$ here is the center of the heteroclinic orbit. 

Note that $v_2$ is a center and $v_1$, $v_3$ are both saddles. We find each of the $v_{x}-v$ panel has a homoclinic or heteroclinic orbits passing through the saddle point, which simplifies the study on the spatially periodic perturbation and helps us to depict the chaos of black holes in phase space. 

Moreover, to research the spatial chaos with perturbation by Melnikov method. Considering $n=1$, $\Omega_{n}$ allows
\begin{equation}
  \Omega_{1}=\begin{pmatrix} 0&1\\-1&0 \end{pmatrix},
\end{equation}
 combining with Eq. (\ref{eq.13}), we can obtain the Melnikov function for these orbits with the above characteristics, which yields
\begin{align}
  M(x_0)=\int_{-\infty}^{+\infty}\left[\frac{v_{x0}(x-x_0)\cos px}{v_0(x-x_0)} \right]dx.
\end{align}

As $\mathbf{z}_0=\left(v_0, v_{x0} \right)^{T}$ is the solution of $\dot{\mathbf{z}}=f(\mathbf{z})$, we could rewrite this function
\begin{align}
  M(x_0)=-L\cos px_0+N\sin px_0,
  \label{MX}
\end{align}
with
\begin{align}
  L & =\int_{-\infty}^{+\infty}\frac{v_{x0}(x)\cos px}{v_0(x)}dx, \\
  N & =\int_{-\infty}^{+\infty}\frac{v_{x0}(x)\sin px}{v_0(x)}dx.
\end{align}

As we can see that the Melnikov function in Eq. (\ref{MX}) has a simple zero at $x_0=\arctan (\frac{L}{N})/p$. Especially, $M(x_0)$ will reach zero when $N=L=0$, and we can obtain $x_0 = k\pi/p$ with $k \in Z$ when $L = 0$ and $N \neq 0$. In PFDM immersed RN-AdS black holes, regardless of the spatial perturbation intensity, the zero point of the Melnikov function causes the thermal chaos which is similar to other charged AdS black holes \cite{Chabab:2018lzf,Mahish:2019tgv,Chen:2019bwt,Dai:2020wny,Wang:2022oop}. The PFDM doesn't affect the existence of zero point for the Melnikov function, it does affect the preconditions $T<T_c$ for generating chaos under a spatial perturbation.   

\begin{figure}[t]
  \centering
  \renewcommand{\figurename}{Fig}
  \subfigure[]{\includegraphics[width=5.4cm]{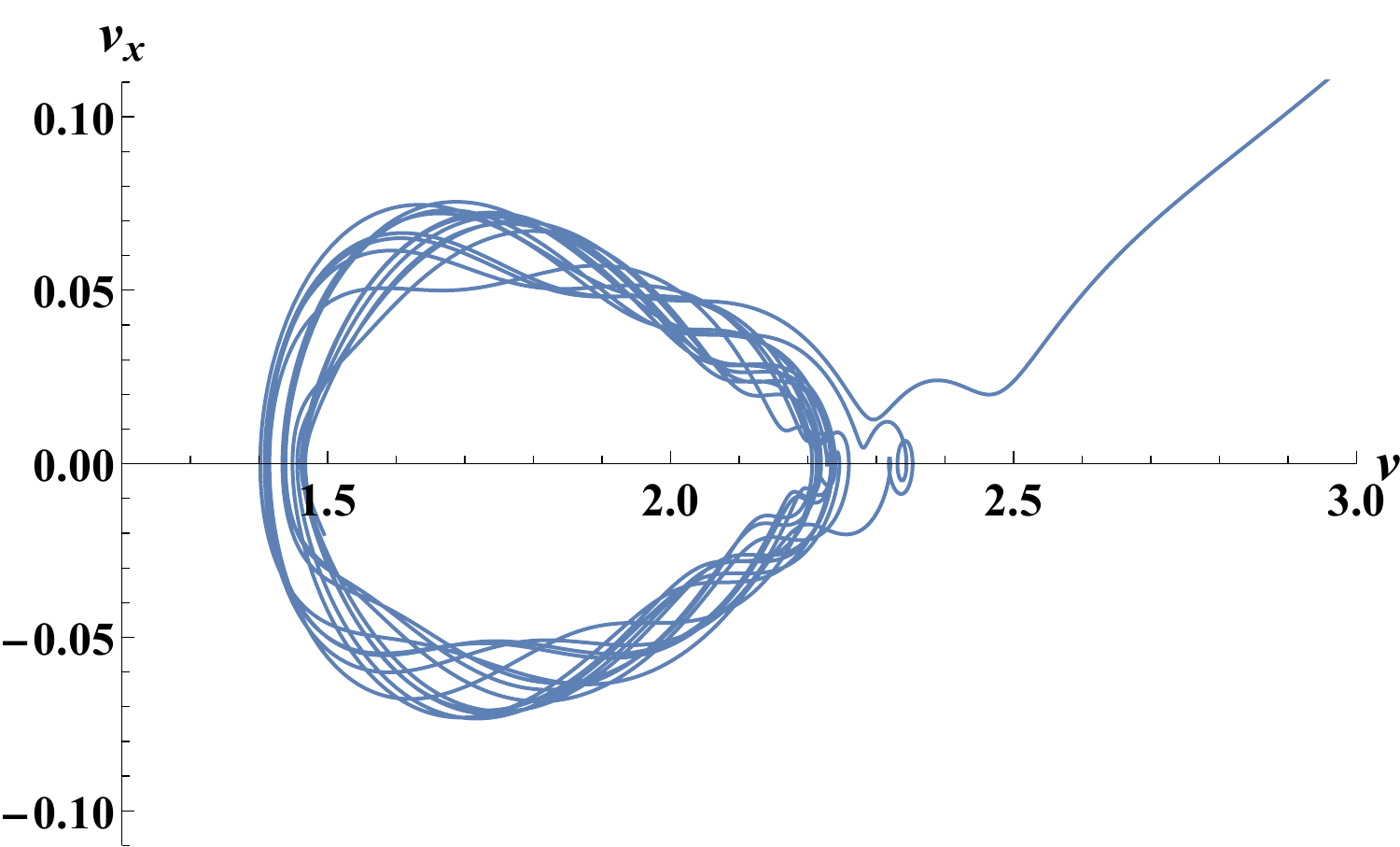}\label{Schaos1}}
  \subfigure[]{\includegraphics[width=5.4cm]{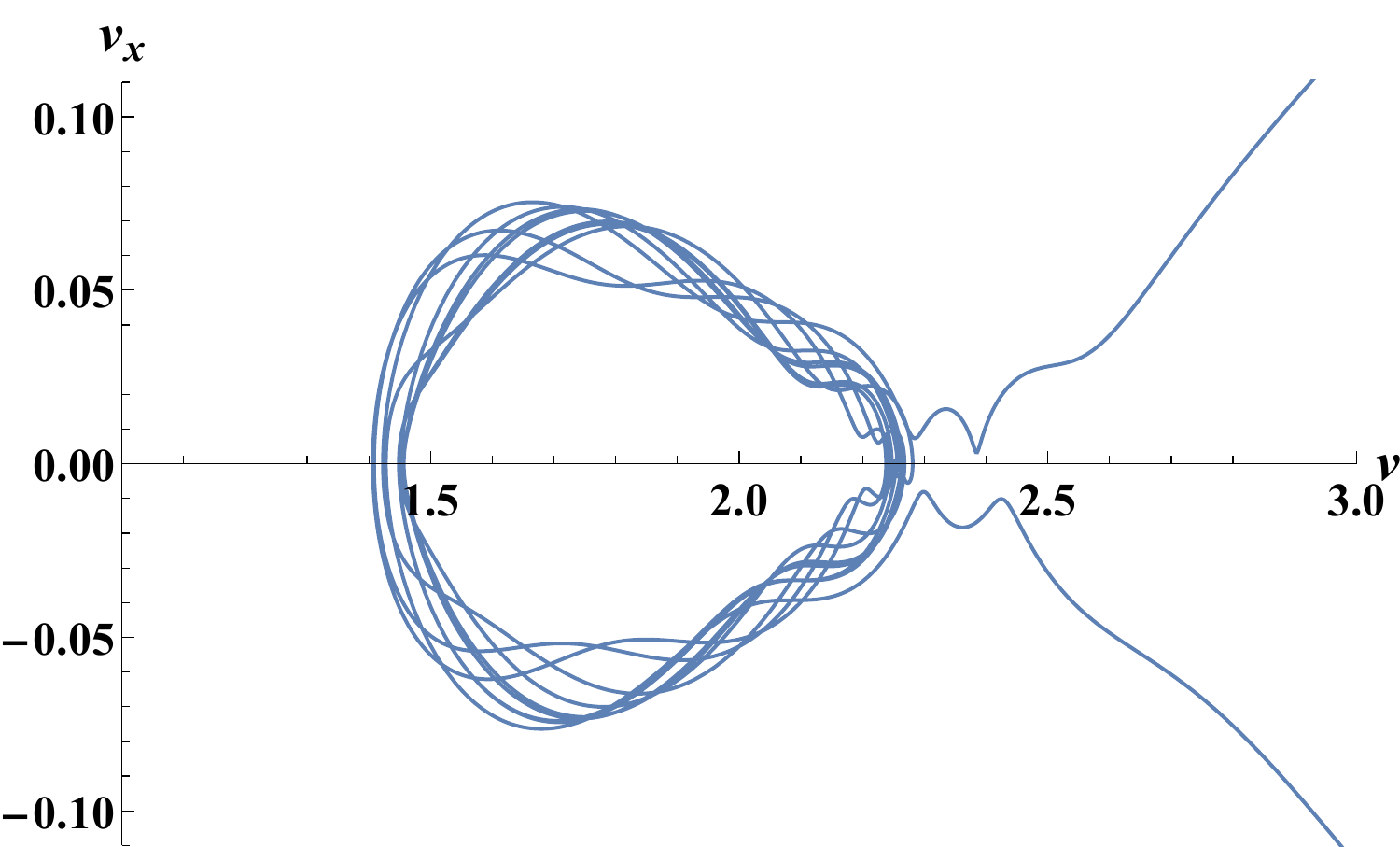}\label{Schaos3}}
  \subfigure[]{\includegraphics[width=5.4cm]{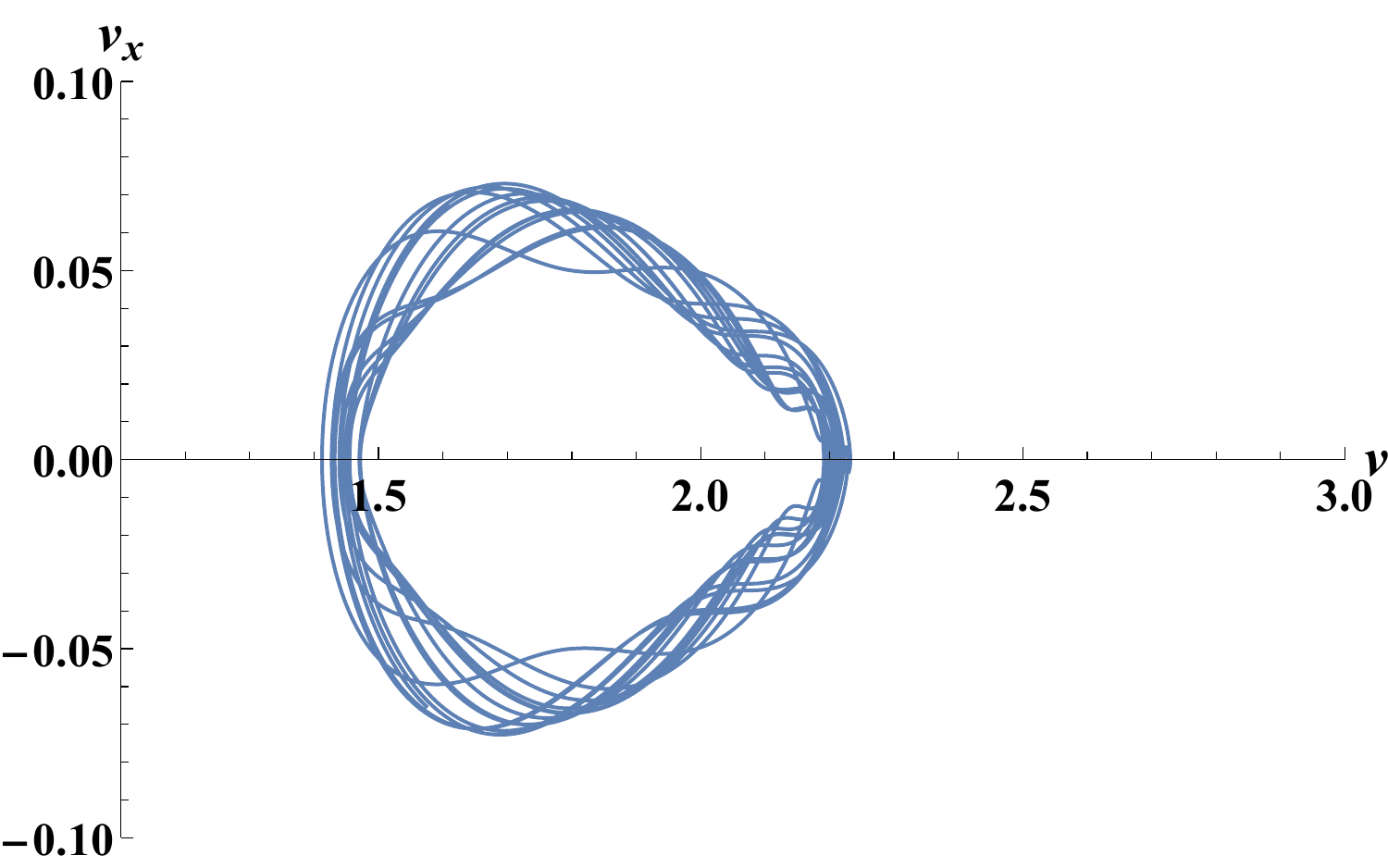}\label{Schaos2}}
  \caption{Phase portraits of the perturbed equation in $v-v_{x}$ plane. The left, the middle and the right panels correspond to the cases with different initial conditions, where we set $\varepsilon=0.002,\, p=0.6$.}
  \label{Schaos}
\end{figure}

Though spatial chaos always occurs in spinodal region, the complex evolution curves near the saddle point will only produce three kinds of chaos according to divergence of evolution image as shown in Fig. \ref{Schaos}. By selecting a closed homoclinic orbit or heteroclinic orbit as the initial configures and plotting the phase portraits of the perturbed equation around the saddle points, we find that the complex and strange shapes in phase planes either diverge in the positive direction in Fig. \ref{Schaos1} or on both sides in Fig. \ref{Schaos3}. When the appropriate initial conditions are selected well, the evolution orbit in Fig. \ref{Schaos2} will be closed but still complex and unpredictable.

\section{Conclusion}

In this work, we first analyzed the thermodynamics of PFDM immersed RN-AdS black holes in the extended phase space and got the relationship of the thermodynamic pressure and the specific volume. We found that the thermodynamic chaos exists only when $T<T_c$ in the spinodal region. To investigate the temporal and spatial chaos, we used the Melnikov method to derive the distance from stable to unstable manifolds in Poincare surfaces. We can obtain the Melnikov function with the dynamical equations and the Hamiltonian for the fluid system. If the zero point of Melnikov function $M(t_{0})$ exists when considering suitable perturbation, the two kinds of manifolds will intersect transversely at $t_0$ which represents the existence of chaos.

After constructing the dynamical equation and adding a weak temporally periodic perturbation, we got the Hamiltonian equations for the fluid system. With the method of Melnikov, analytical solution of a typically homoclinic orbit was obtained for unperturbed system. When the temporal perturbation amplitude larger than the critical value $\gamma_c$ the homoclinic orbit's evolutionary process became extremely complex and unpredictable, which is similar to the cases of some charged AdS black holes \cite{Chabab:2018lzf, Mahish:2019tgv, Chen:2019bwt, Wang:2022oop, Dai:2020wny}. Moreover, the dependence of $\gamma_c$ on the black holes charge $Q$ and the PFDM parameter $\lambda$ has been investigated with initial temperature $T_0$ proportional to the critical value $T_c$ or has a fixed value. For $T_0=0.9T_c$, when considering PFDM we found that it is difficult to produce temporal chaos for charged black holes, but easy for uncharged AdS black hole. When $T_0$ fixed, it would bring an lower bound for $\lambda$ or upper bound for $Q$ due to the restriction of $T_{0}<T_c$.

Similar to the research of temporal chaos, we obtained the ambient pressure $B$ and got dynamical equation with a tiny spatially thermal periodic perturbations. According to the value of $B$ and phase transition pressure $P_0$, three different types of portraits were plotted in the phase plane in Figs. \ref{f7}, \ref{f8} and \ref{f9}. In all three cases, there are one center $v_2$ and two saddles $v_1$, $v_3$ which represent the equilibrium point of the homoclinic or heteroclinic orbits. Then, selecting a closed homoclinic orbit or heteroclinic orbit as the initial configures, we plotted the phase portraits of the perturbed equation around the saddle points to check the local equilibrium in Fig. \ref{Schaos}. Most of the time, the complex and unpredictable shapes in phase planes diverge either in the one direction or on both sides. In particular, when the appropriate initial conditions are selected, the evolution orbit will have no divergence. By solving the Melnikov function and analyzing the solution, it shows that the spatial chaos appears regardless of perturbed amplitude which is similar to that in the case of VdW fluid system and other black holes \cite{Chabab:2018lzf,Mahish:2019tgv,Chen:2019bwt,Wang:2022oop,Dai:2020wny}.

We have studied the chaos for RN-AdS black holes immersed in PFDM under temporally/spatially periodic perturbations in the extended phase space by analogy with the VdW fluid. It would be interesting to investigate these thermodynamic chaos for black holes with PFDM. We hope this work could inspire us to understand the thermodynamic properties of black holes with a thermal fluctuation in a more reasonable way.

\begin{acknowledgments}
  We are grateful to Aoyun He, Wei Hong and Lingquan Ma for useful discussions and valuable comments. This work is supported by NSFC (Grant No. 12175212 and 12147207).
\end{acknowledgments}

\end{document}